\documentstyle[titlepage,fleqn,12pt]{article}
\begin{document}
\begin{titlepage}
\title{Solitons and $1/f$ noise in molecular chains}
\author{
\\
\\
H. Rosu\\
{\it Instituto de Fisica, Universidad de Guanajuato, Le\'on, M\'exico}
\\
\\
\hspace{0.01cm}E. Canessa\\
{\it ICTP-International Centre for Theoretical Physics, Trieste, Italy}
}

\date{}
{\baselineskip=25pt

\begin{abstract}

Davydov's model of solitons in $\alpha$-helix protein chains is shown
to display features of self-organized criticality (SOC), {\em i.e.},
power law behaviour of correlations in space and $1/f$ noise, as a
consequence of considering {\em random} peptide group displacements from
their (periodic) equilibrium positions along a chain.  This may shed light
on a basic mechanism leading to obtain flicker noise in $\alpha$-helix
protein chains and to predict a SOC regime in biomolecular structures
from first principles.  We believe our treatment of $1/f$ noise to be of
some relevance to recent findings due to Voss on DNA [{\em Phys.
Rev. Lett.} {\bf 68}, 3805 (1992)].

\vskip 2cm


\end{abstract}

    }
\maketitle
\end{titlepage}
\baselineskip=25pt
\parskip=0pt

The concept of solitons found a fascinating novel application into
biological phenomena since Davydov \cite{Dav73} introduced a cubic non-linear
Schr\"{o}dinger equation to describe energy transport in molecular chains such
as $\alpha$-helices.  The topological and dynamical stability of Davydov
solitons -{\em i.e.}, conservation of form and velocity after interaction,
respectively- is related to the spontaneous (local) symmetry breaking of
the protein molecules.  A dynamical balance between the dispersion due to the
resonance interaction of intrapeptide dipole vibrations and the nonlinearity
of the interaction of such vibrations with the (time-dependent) local
displacements from the equilibrium positions of the peptide groups -say,
$\beta$- is believed to play a crucial role for the transport of energy
released in the hydrolisis of the {\em adenosine triphosphate} (ATP)
molecule \cite{Hoc71}.

Because of this, Davydov solitons -as well as alternative soliton models
to account for nonlinearity in quasi 1D biomolecular chains (see, {\em e.g.},
\cite{Mut90,Tak89,Yom85})- are an active research field.  Recent progress has
been directed to relate excitations in Davydov systems to Bose-Fr\"{o}hlich
condensation phenomena \cite{Ris92,Bol91} and to include more
appropriately quantum and temperature effects on the lifetime of the soliton
\cite{Cru92,Sch92}.  Even though a large amount of theoretical work has been
accumulated in the past two decades following these Davydov's pioneering ideas,
there is still a number of open problems.   Firstly, there is as yet no
experimental verification of Davydov solitons \cite{Fan90,Sco91}.  Indeed
Davydov's model might look rather unrealistic \cite{Ris92}, but from the
standpoint of physics it is not unreasonable to study it as an effective way
towards achieving better understanding of complex molecular systems.
Secondly, to our knowledge, the case of having {\em random} peptide group
displacements $\beta$ due to {\em local disorder} has not been fully
considered.

To this end the recent proposal by Voss \cite{Vos92} (see also
\cite{Pen92,Kan92}) that long-range correlations and $1/f$ noise
can be detected in DNA sequences when viewed as random processes,
gives one inspiration for carrying out an investigation on random $\beta$.
The idea that nucleotide bases in strands of DNA may be correlated over
several thousands of base-positions, opens the way to explore how, and
to which extent, solitons might be useful to
gain physical insight into Voss's relevant finding
on DNA.  It is tempting to analyse this phenomenon from a first
principle formalism based on soliton physics and, from this, to understand
that class of non-linear dynamical systems which drive themselves into a
statistically stationary critical state, the so-called self-organized
criticality (SOC) \cite{Bak87}, with no intrinsic length or time scale,
where the systems exhibit power law (fractal) behaviour and generate
flicker noise \cite{Note1}.

In this work we shall make a small step towards such a theory to show that
Davydov solitons in quasi-1D $\alpha$-helix chains at $0 \;K$, if
they exist, might display features of SOC as a consequence of assuming
{\em random} peptide group displacements from their
(periodic) equilibrium positions along a chain.  This suggests a possible
physical mechanism to understand SOC from an {\em ab-initio} basis in terms
of interactions.  In addition our work complement recent ideas put forward
by one of us \cite{Can93}, regarding an analytical continuous probability
theory for SOC, in that we propose here that random peptide group
displacements might be one possible
mechanism to generate $1/f$ noise in macromolecular chains.

We start considering the simplest Davydov's protein molecule,
{\em i.e.} $\alpha =1$ \cite{Dav73}, which we briefly describe
below for completeness.
The basic idea is that the amide-$I$ vibrational energy is coupled
through interaction with acoustic phonons.
The molecular chain has N ($\gg 1$) molecular units of
mass $M$ ($\sim 114 \; a.m.u.$) placed at positions: $x_{n}= nR + u_{n}$,
where $n$ is an index that counts unit cells (in the hydrogen-bonded
direction), $u_{n} (\ll R) $ are small displacements from the equilibrium
positions $nR$ caused by internal molecular motion.
The molecular groups are taken as the peptide subunits of $\alpha$-helix
protein polymers and $R$ as the length of the amide's hydrogen bond
($\sim 3\AA$).

The Hamiltonian operator $H$ for the collective
degrees of freedom resulting from the interaction of intramolecular amide-$I$
modes, ($C=O$ stretching), and the lattice motion of such a chain is
\begin{equation}\label{eq:d1}
H= H_{vib} +H_{ph}+ H_{int}   \;\;\; .
\end{equation}
In the above,
\begin{equation}
H_{vib}= \sum _{n}\{ \epsilon _{0}B_{n}^{\dagger}B_{n}-
     J(B_{n}^{\dagger}B_{n+1}+B_{n+1}^{\dagger}B_{n}) \}  \;\;\; ,
\end{equation}
where $B_{n}^{\dagger}$ and $B_{n}$ are boson creation and annihilation
operators for the vibrational excitation at the $n$th site associated with
the amide-$I$ dipolar oscillator -having the quantum energy
$\epsilon _{0}\sim 0.205\; eV$.
The dipolar resonant interaction is only considered between the nearest
neighbor molecules, {\em i.e.}, $J= 2d^{2}/R^{3}$
($=9.67\times 10^{-4}\; eV$ from infrared spectra), where $d$
($= 0.29\; D$) is the dipolar electric moment aligned along the $C=O$ bond.

The second term of Eq.(\ref{eq:d1}) describes the longitudinal
harmonic oscillations (phonons) of the chain, which in second quantized form
is written as a sum over the momentum normal modes $q$, namely
\begin{equation}
H_{ph}= \sum _{q} \hbar \Omega _{q} (b_{q}^{\dagger}b_{q} +\frac{1}{2})
                 \;\;\; ,
\end{equation}
where $b_{q}^{\dagger}$ and $b_{q}$ are phonon creation and annihilation
operators.  $\Omega _{q}$ is given by the dispersion equation
$\Omega _{q}^{2}=4 (\nu _{a}^{2}/R^{2})\sin ^{2}(\frac{1}{2} qR)$
with $\nu _{a}= R\sqrt{w/M}$ and $w$ an elasticity coefficient
($\sim 76\; N/m$).

The last term in the collective Davydov Hamiltonian is the nonlinear
interaction between the vibrational dof's and the phonon dof's, namely
\begin{equation}
H_{int}=\frac{1}{\sqrt{N}}\sum_{q,n} \chi (q)e^{iqnR}
  B_{n}^{\dagger}B_{n}(b_{q}+b_{-q}^{\dagger})  \;\;\; ,
\end{equation}
where
$\chi (q) =\chi ^{*} (-q) =i\chi(\frac{\hbar}{2M\Omega _{q}})^{1/2} \sin (qR)$.
The physical meaning of the nonlinear coupling parameter $\chi$ is
to characterize change effects of the amide-$I$ bond energy
per some unit extension of the hydrogen bond.  This parameter
is expressed as $\chi =\partial \epsilon _{0}/\partial R$,
whose numerical value lies in the interval:
$\chi =3-6.2\times 10^{-11} \; N$.

For the wavefunction of the above Schr\"{o}dinger equation,
Davydov wrote
\begin{equation}\label{eq:d2}
||\psi _{_{D}}(t)\rangle \rangle=
\sum_{n} \alpha_{n}(t) \exp ({\sigma (t)}) B_{n}^{\dagger}
            \; ||0\rangle \rangle  \;\;\; ,
\end{equation}
where
$||0\rangle \rangle$ is a generalized vacuum state in the collective
space of vibrational-phonon dof's and $\alpha _{n}(t)$
are normalized functions.
The quantum-mechanical phase $\sigma (t)$ is written as
$\sigma (t) =-\frac{i}{\hbar}
\sum_{q}[\beta _{qn}(t)b_{q}^{\dagger}-\beta _{qn}^{*} (t) b_{q}]$,
which is termed the $D_{1}$ ansatz.  It is a superposition of
tensor products of single-exciton states and coherent phonon states
\cite{Bro86}.  There also exists a classical displaced oscillator ansatz
(or $D_{2}$) as well as a modified $m-D_{1}$ state \cite{Bro89}.

For the complex functions $\alpha _{n}(t)$ and the
real functions $\beta_{qn} (t)$ of Eq.(\ref{eq:d2}) -which characterize
the vibrational state and the displacement from equilibrium of a single
molecular unit at site $n$, respectively-
Davydov obtained a system of two coupled discrete-differential equations
expressing their `classical' Hamiltonian evolution.
In the long wavelength approximation such discrete functions are
replaced by two continuous limits $\alpha (\zeta )$ and $\beta (\zeta )$,
where $\zeta\equiv nR-v_{s}t$, subject to the constraint that
the propagation velocity $v_{s}$ is constant ({\em i.e.}, stationary
propagation).  A further approximation is to take $nR$ as any point $x$
along the chain axis.  Besides $\beta$, the function
$\rho (\zeta)=-R\frac{\partial \beta (\zeta)}{\partial \zeta}$ is also
defined to characterize the (infinitesimal) difference of displacements
of nearest neighbour molecules within the chain and
modulation of the chain due to vibrational solitons.

Solutions of the energy transport within Davydov's
Hamiltonian treatment in the continuous and subsonic regime
({\em i.e.}, $s^{2}=v_{s}^{2}/v_{a}^{2}\ll 1$) are as follows:
\begin{eqnarray}
\alpha(\zeta ) & = & \sqrt{\mu /2}
e^{[\frac{i}{\hbar}[\frac{\hbar ^{2}v_{s}x}{2JR^{2}}-E_{s}t]]}
cosh ^{-1}(\frac{\mu}{R}\zeta)  \;\;\; , \label{eq:dav1} \\
\rho (\zeta) &  = &  \frac{\chi}{w(1-s^{2})}
    sech ^{-2} (\frac{\mu}{R}\zeta) \label{eq:dav2}\\
\beta (\zeta) &  =   &  \frac{\chi}{w(1-s^{2})} (1-tanh (Q\zeta ))  \;\;\; ,
               \label{eq:dav3}
\end{eqnarray}
where $Q=\frac{MR\chi ^{2}}{2w\hbar ^{2} (1-s^{2})}$.
The quantities $\mu$, $E_{s}$, and $s$ are respectively given by
$\mu=\chi ^{2}/Jw(1-s^{2})$,
$E_{s}=\epsilon _{0}-2J+\frac{\hbar ^{2}v_{s}^{2}}{4JR^{2}}-J\mu^{2}/3$,
and $s=v_{s}/v_{a}$.

In the above $\alpha(\zeta )$ is essentially the Davydov vibrational soliton,
which is a quasi-localized structure with size of the order $R/\mu$,
propagating with velocity $v_{s}$ and transfering vibrational energy
$\epsilon _{0}$.  Furthermore, $\rho (\zeta)$ is the hydrogen bond soliton
whereas $\beta (\zeta)$ is a kink soliton of the displacements of the
peptide groups from their (periodic) equilibrium positions.
During their (constant-$v_{s}$) movement, the three solitons (usually denoted
as $S_{1}$, $R_{1}$ and $K_{1}$, respectively)
strongly influence each other in a way which is still an open
problem.  Of special interest to us here is the $K_{1}$ solution, which
might be considered as a spatial continuous representation of
D'Alambert type ({\em i.e.}, of the form $\beta (x-v_{s}t)$)
of a product of coherent states, that is needed to obtain the
nonlinear Schr\"{o}dinger dynamics \cite{Bliw86}).

Let us include next, to the above Davydov model, a special type of
{\em random} disorder that may result from several sources, such
as radiation and others, and that fluctuates in time.
We shall come back to this latter on.
In the $\zeta$-coordinate frame, the central position of
the kink $K_{1}$ can be fixed as the spatial origin and small time
excursions of its origin will be allowed.  Then it is straightforward to
estimate time correlations via the (dimensionless) noise power spectrum
of the temporal evolution of such fluctuations by characterizing random
processes in $\beta$ of Eq.(\ref{eq:dav3}) at the time scale
$0<t<\tau\approx 1/f$ by \cite{Vos92}
\begin{equation}
S_{\beta}(f) \propto \frac{1}{\tau}  \mid \int_{0}^{\tau} \beta (\zeta)
       e^{2\pi i ft} \; dt  \mid^{2}    \;\;\; ,
\end{equation}
where $\tau \rightarrow \infty$ and
$\zeta (x,t)$ is the random variable.  Figure 1 shows the results
obtained for the spectral density $S_{\beta}$ when using
$x\rightarrow x_{o}\pm \Delta x$, such that
we set $x_{o}=0$ and choose $\Delta x$ to vary randomly between
$\pm 1$ for simplicity.  These results have been
computed using a standard fast fourier transform algorithm. On
averaging over thousand ensambles, $2^{13}$ (unit-)time steps within the
interval $0<t<10^{4}$ are found to suffice in order to achieve less than
$5\%$ of error in calculations.  As revealed from this figure the noise
power spectrum of the Davydov $\beta$ displacement, when assumed to be a
function of random $\zeta$, shows a clear manifestation of $1/f$ noise.
We believe this treatment of flickler noise to be somehow of
relevance to the findings due to Voss on DNA \cite{Vos92}. In particular,
we can superpose a peaked structure in the $1/f$ power spectrum of $\beta$
by periodically resetting $\Delta x$ to naught as shown in the high-frequency
region of Fig.1.

{}From these results we can see that the transition to a
steady state, is strongly related to the peptide group fluctuations from
their initial ({\em i.e.}, $t_{o}=0$, $x_{o}=0$) conditions.
According to such complex dynamics, the physics behind this phenomenon
-{\em i.e.}, existence of time correlations-
is due to the random disorder we have introduced and whose possible cause
will be analysed below.

Let us first also investigate the space domain by
fixing the time scale ({\em i.e.}, by assuming now static solutions).
We consider once more the random variable $\zeta$ to characterize random
disorder of the peptide group displacements from their equilibrium positions
with a density probability $\phi$ proportional to $\beta (\zeta)$ \cite{Can93}.
Since the uniform {\em probability distribution function} of having
random events can be written as
\begin{equation}\label{eq:can2}
{\cal G}(\zeta_{2} )-{\cal G}(\zeta_{1}) ={\cal P}
     \{ \zeta_{1}< {\bf \zeta} \leq \zeta_{2} \} \approx
     \int_{\zeta_{1}}^{\zeta_{2}}  \phi (\zeta ')
       \; d\zeta '    \;\;\; ,
\end{equation}
where $\{ \}$ indicates the function interval.
Then, this integral over the limits
$\zeta_{2} \equiv \lambda_{1} \zeta_{o} \geq \zeta + \lambda_{2}
\zeta_{o}\equiv \zeta_{1}$ gives
\begin{equation}\label{eq:cca1}
 {\cal G}(\lambda_{1} \zeta_{o}) - {\cal G}(\zeta + \lambda_{2} \zeta_{o})
    = \int_{\zeta  + \lambda_{2}
            \zeta_{o}}^{\lambda_{1} \zeta_{o}} \beta (\zeta ') \; d\zeta '
      \equiv -\tau (\zeta )  \;\;\; ,
\end{equation}
with $\beta$ given in Eq.(\ref{eq:dav3}), ($s^{2}\ll 1$),
and $\zeta_{o}$ being a constant.
The minus implies that the functions ${\cal G}$ are here assumed to satisfy
the condition
${\cal G}(\zeta + \lambda_{2} \zeta_{o}) > {\cal G}(\lambda_{1}\zeta_{o})$
for $\zeta \neq 0$, which do not need to be defined, whereas the
free parameters $\lambda_{i}$ ($i$=1,2) will restrict the range of $\zeta $.

The afore-mentioned integration limits lead to the condition
$\lambda_{2} -\lambda_{1} +  \zeta  \leq 0$ as discussed in our previous
work \cite{Can93}.
If $\lambda_{2}\neq \lambda_{1}$, then we get (in terms of $\tau (0)$)
\begin{eqnarray}\label{eq:w1}
\tau (\zeta^{*} ) & \approx & (1+\frac{\zeta^{*}}{\lambda_{2} -\lambda_{1}})
    \{ \tau (0)+ \zeta_{o} \ln cosh \lambda_{2}\}  \nonumber \\
        &   & \hspace{2.2cm} - \zeta_{o}\{ \frac{\zeta^{*} }
      {\lambda_{2} -\lambda_{1}}\ln cosh \lambda_{1}
      + \ln cosh (\lambda_{2}+\zeta^{*})  \}   \;\;\; .
\end{eqnarray}
where $\zeta^{*}=\zeta/\zeta_{o}$,
$\tau (0)=\zeta_{o}(\lambda_{2}-\lambda_{1})(1+\Gamma_{\lambda})$ and
$\Gamma_{\lambda}=(\ln cosh \lambda_{1}-\ln cosh
\lambda_{2})/(\lambda_{2}-\lambda_{1})$.

In Fig.2 we show the dependence of the normalized probability
distribution function $\tau$ on the reduced variable $\zeta ^{*}$
for values of $\lambda_{1}=3$, $\lambda_{2}=-8$ and $\tau (0)=1$
which, in turn, determine the value of $\zeta_{o}$.
In our calculations, {\em i.e.} using Eqs.(\ref{eq:dav3}) and Eq.(\ref{eq:w1}),
we have also reduced
$\frac{\chi}{w(1-s^{2})}\propto 1$ and $Q\zeta_{o}\propto 1$ for simplicity.
The choice of $\tau(0)$ allows to normalize $\tau (\zeta^{*})$
and to mimic features of SOC, namely a power-law behaviour in space
correlations (within the range $0\le \zeta^{*}\le 8$), provided $\zeta^{*}$
is associated with the $log$-function of a measured random event.  In fact,
we have that $\tau (\zeta^{*}\rightarrow 0) \rightarrow 1$ and
$\tau (\lambda_{1}-\lambda_{2})\equiv 0$.  Whereas if
$\lambda_{2}+\zeta^{*}\approx 0$ then
$\ln cosh (\lambda_{2}+\zeta^{*})\approx 0$, hence $\tau$
of Eq.(\ref{eq:w1}) depends linearly on $\zeta^{*}$ for values
$\zeta^{*}\leq -\lambda_{2}$.

To see more clearly possible power-law features in the behaviour of
$\tau$ over an extensive range of values of $\zeta^{*}\leq 8$,
we calculate the linear derivative of $\tau (\zeta^{*})$
which is also plotted in Fig.2 by dotted lines.
For the smallest displayed values of $\zeta^{*}$ this function
converges to a constant negative value, revealing in this way the constant
nature of the negative slopes in the ($\tau$-$\zeta^{*}$) curves.
In view of these features of the derivative of $\tau$, the second
derivative has also been included in Fig.2.  It presents a sharp peak around
the inflection point of $\frac{\partial \tau}{\partial \zeta^{*}}$, thus
indicating the range of validity of such a power-law behaviour.  The cutoff
in the $\zeta^{*}$-axis for the $\tau$ curve is related to the system
size or integration limit ({\em i.e.}, $\lambda_{1}-\lambda_{2}\approx 11$).
Accordingly, it reflects the range of long-range correlations in the
space domain.
In view of these results we can interpret $\beta$ as a nonequilibrium order
parameter of a transition from the $D_{1}$ phase of the chain to the
Fr\"{o}hlich phase \cite{Fro68}.  That is, a phase transition from the
dynamical balance between the intramolecular excitations (and their exchange)
and the longitudinal excitations of the linear Davydov chain, to the
Fr\"{o}hlich nonthermal excitations of longitudinal polarization
modes arising in a far-from-equilibrium regime supported by the flow of energy.

We focus now on some possible effects that have been usually
neglected when taking a continuous limit within the framework of
the Davydov theory but which, in our opinion, may become important
as being the source for generating randomness.
These effects include radiation, discreteness-chaos, and disorder.
Radiation effects have been first discovered in the context of lattice
topological solitons (dislocations), and subsequently proved to exist also
in the case of dynamical solitons \cite{Pey86}.
In fact, subsonic kinks in a monoatomic chain permanently radiate small
amplitude oscillations.  Besides this, subsonic kinks (as well as supersonic
ones) in diatomic chains lose energy in this process \cite{Pey86}.
Chaotic effects, due to lattice discreteness have been discussed
in Ref.\cite{Tak85} within a vibron model, with on-site
potentials, which is very similar to the Davydov model.
Such an effect can be understood as a perturbation of an
integrable system.  It is a chaotic effect that might imply a random
spatial arrangement of stationary solitons.  The effect of
disorder, on the other hand, is a very important feature of many nonlinear
chains \cite{Phi87}.
Considering the 20 different natural amino acids in real peptide chains,
each of different mass, one may think of a sequence of random masses to
have important effects on soliton propagation.
This has been observed by F\"{o}rner \cite{For91} by investigating
sequences of masses, spring constants, nonlinear coupling constants, heat
bath, and disorder in the dipole coupling.  Another crucial effect
to consider when applying the continuous limit is the impurity disorder in
which various types of kink-impurity interactions may be possible
\cite{Fei92}.  We believe all of these phenonomena to generate (a sort of
intrinsic) randomness
for the displacements $\beta$ along a single molecular chain ({\em i.e.},
$\alpha =1$), which leads to obtain SOC features as we have discussed
in this work.  We add that the present ideas may be easily extended to
chains with $\alpha >1$, since
the soliton solutions of the Davydov Hamiltonian -{\em i.e.},
$S_{1}$, $R_{1}$, $K_{1}$- keep their form unchanged \cite{Dav73}.

We have thus tried to interpret, within the simplest Davydov soliton
theory, randomness by combining features of both discreteness
and disorder effects.  These considerations may be useful to
understand long-range correlations in biological systems \cite{Man92}
and flicker noise, possibly including DNA \cite{Vos92}.  We have predicted
a SOC dynamical regime in biomolecular systems from first principles.
Indeed we have been able to derive such a SOC regime as a
consequence of random $\beta$ which has been interpreted in terms of a
nonequilibrium phase transition between the $D_{1}$ soliton state of the
chain and a probable Fr\"{o}hlich- condensed phase.  The present SOC regime
might be seen as a proof of the soliton stability against random
disorder.  In turn, this may be an indication of the nonequilibrium
nucleation of Fr\"{o}hlich domains along a single chain.  Of course, to
characterize randomness of the peptide group displacements
with a density probability $\phi \propto \beta (\zeta)$
may be seen heuristic.  But, in view of the results obtained, this can
be considered as one of the simplest reasonable ways to
relate local properties to macroscopic behaviour.  To this end, note
that a relation of this kind has also been used within the context of
pattern formation \cite{Can92}.

\vspace{0.5cm}

The authors thank Dr.J. Langowski (EMBL-Grenoble) for a valuable discussion
and Prof. Yu Lu (ICTP-Trieste) for encouragement.

\newpage

\section*{Figure captions}

\begin{itemize}

\item {\bf Fig.1}: Noise power spectrum
for the temporal evolution of the random fluctuations
in $\beta (\zeta^{*})$ including periodical resetting.

\item {\bf Fig.2}: (Full line) probability distribution function $\tau$
of having random events {\em vs.} reduced variable $\zeta^{*}$
using $\lambda_{1}>\lambda_{2}$, such that $\lambda_{2}<0$, and
$\tau (0)=1$.
(Dotted lines) first and (smallest dotted lines) second derivative
of $\tau$ with respect to $\zeta^{*}$.

\end{itemize}


\newpage

\begin{thebibliography}{99}


\bibitem{Dav73} A.S. Davydov, J. Theor. Biol. {\bf 38}, 559 (1973);
see also {\em 'Solitons in Molecular Systems'}, Reidel (1985).
\bibitem{Hoc71} F.L. Hock, {\em 'Energy Transformations in Mammals:
Regularity Mechanisms'}, Saunders (1971) pp.1-5.
\bibitem{Mut90} V. Muto, A.C. Scott and P.L. Christiansen,
Physica D {\bf 44}, 75 (1990).
\bibitem{Tak89} S. Takeno, Prog. Theor. Phys. {\bf 69}, 1689 (1983);
J. Phys. Soc. Japan {\bf 59}, 3127 (1990).
\bibitem{Yom85} S. Yomosa, Phys. Rev. A {\bf 32}, 1752 (1985).
\bibitem{Ris92} Lj.M. Ristowski, Z. Nestorovi\v c and G.S. Davidovi\v c,
Z. Phys.B {\bf 88}, 145 (1992).
\bibitem{Bol91} H. Bolterauer, J.A. Tuszy\v nski and M.V. Satari\v c,
Phys. Rev. A {\bf 44}, 1366 (1991).
\bibitem{Cru92} L. Cruzeiro-Hansson, Phys. Rev. A {\bf 45}, 4111 (1992).
\bibitem{Sch92} J.W. Schweitzer, Phys. Rev. A {\bf 45}, 8914 (1992).
\bibitem{Fan90} W. Fann {\em et al}., Phys. Rev. Lett. {\bf 64}, 607 (1990).
\bibitem{Sco91} A.C. Scott, Physica D {\bf 51}, 333 (1991).
\bibitem{Vos92} R.F. Voss, Phys. Rev. Lett. {\bf 68}, 3805 (1992).
\bibitem{Pen92} C.-K. Peng {\em et al.}, Nature (London) {\bf 356},
168 (1992).
\bibitem{Kan92} W. Li and K. Kaneko, Europhys. Lett. {\bf 17}, 655 (1992).
\bibitem{Bak87} P. Bak, C. Tang and K. Weisenfeld, Phys. Rev. Lett
{\bf 59}, 381 (1987).
\bibitem{Note1} It has been found that discrete (sandpile) models for SOC
result in $1/f^{2}$ whereas their continuos couterparts may exhibit
$1/f$ instead.  Accordingly, we obtain here the continuos limiting
behaviour.  For a discussion see H.F. Chau and K.S. Cheng, Phys. Rev. A
{\bf 46} R2981 (1992).
\bibitem{Can93} E. Canessa, Phys. Rev. E {\bf 47}, R5 (1993).
\bibitem{Bro86} D.W. Brown, B.J. West and K. Lindenberg
Phys. Rev. A {\bf 33}, 4104 (1986); {\em ibid} {\bf 33}, 4110 (1986).
\bibitem{Bro89} D.W. Brown and Z. Ivic, Phys. Rev. B {\bf 40}, 9876
(1989).
\bibitem{Bliw86} D.W. Brown, K. Lindenberg and B.J. West,
Phys. Rev. Lett. {\bf 57}, 2341 (1986).
\bibitem{Fro68} H. Fr\"{o}hlich, Int. J. Quantum Chem. {\bf 2},
641 (1968).
\bibitem{Pey86} M. Peyrard, S. Pnevmatikos and N. Flytzanis,
Physica D {\bf 19}, 268 (1986).
\bibitem{Tak85} S. Takeno, Prog. Teor. Phys. {\bf 73}, 853 (1985).
\bibitem{Phi87} S.R. Phillipot {\em et al.}, Phys. Rev. B {\bf 35},
7533 (1987).
\bibitem{For91} W. F\"{o}rner, Phys. Rev. A {\bf 44}, 2694 (1991) and
references therein.
\bibitem{Fei92} Z. Fei, Y.S. Kivshar and L. V\'{a}zquez,
Phys. Rev. A {\bf 46}, 5214 (1992); {\em Ibid} {\bf 46}, 6019 (1992).
\bibitem{Man92} G.S. Mani, J. Theor. Biol. {\bf 158}, 447 (1992).
\bibitem{Can92} W. Wei and E. Canessa, Phys. Rev. E {\bf 47}, 1243 (1993).

\end{thebibliography}
\end{document}